\documentclass[copyright,creativecommons]{eptcs}
\providecommand{\event}{SR 2014} 
\usepackage[utf8]{inputenc}
\usepackage{amssymb,amsmath,stmaryrd}

\usepackage{amsthm}
\newtheorem{theorem}{Theorem}
\theoremstyle{remark}
\newtheorem{exa}[theorem]{Example}
\newenvironment{example}{\begin{exa}}{\hfill$\lhd$\end{exa}}

\usepackage{mathtools}
\usepackage{cases}
\usepackage{array}
\usepackage[table]{xcolor} 
\usepackage{tikz}
\usetikzlibrary{arrows,shapes,backgrounds,decorations.pathreplacing}
\usepackage{mathptmx}
\usepackage{multirow}

\usepackage{float}
\usepackage{dsfont}
\DeclareSymbolFont{cmmathcal}{OMS}{cmsy}{m}{n}
\DeclareSymbolFontAlphabet{\mathcal}{cmmathcal}
\usepackage{algorithm}
\usepackage[noend]{algorithmic}
\usepackage{enumerate}
\usepackage{wrapfig}

\newcommand{\state}{\ensuremath{s} }

\newcommand{\game}{\ensuremath{G} }
\newcommand{\reduc}{\ensuremath{\downharpoonright} }

\newcommand{\thresholdWC}{\ensuremath{\mu} }
\newcommand{\thresholdExp}{\ensuremath{\nu} }

\newcommand{\integ}{\ensuremath{\mathbb{Z}} }
\newcommand{\nat}{\ensuremath{\mathbb{N}} }

\newcommand{\player}{\ensuremath{\mathcal{P}} }
\newcommand{\playerOne}{\ensuremath{\mathcal{P}_{1}} }
\newcommand{\playerTwo}{\ensuremath{\mathcal{P}_{2}} }

\newcommand{\stepsWC}{\ensuremath{L} }
\newcommand{\stepsExp}{\ensuremath{K} }

\newcommand{\cmbSum}{\ensuremath{\mathsf{Sum}} }
\newcommand{\typeA}{\ensuremath{\textit{(a)}} }
\newcommand{\typeB}{\ensuremath{\textit{(b)}} }

\newcommand{\NPinter}{\ensuremath{\text{NP} \cap \text{coNP}}}
\newcommand{\PTIME}{\ensuremath{\text{P}}}
\newcommand{\NPTIME}{\ensuremath{\text{NP}}}

\newcommand{\BWC}{\text{BWC}}

\newcommand{\gameTS}{\ensuremath{\game_{\thresholdWC}} }
\newcommand{\tsFailSymbol}{\ensuremath{\top} }

\floatstyle{plain}
\newfloat{myalgo}{tbhp}{mya}
{\begin{myalgo}[#1]
\centering
\begin{minipage}{#2}
\begin{algorithm}[H]}%
{\end{algorithm}
\end{minipage}
\end{myalgo}}

\title{Expectations or Guarantees? I Want It All!\\A crossroad between games and MDPs\thanks{Work partially supported by European project CASSTING (FP7-ICT-601148). Filiot and Randour are respectively F.R.S.-FNRS research associate and research fellow. Raskin is supported by ERC Starting Grant inVEST (279499).}}
\author{V\'eronique Bruy\`ere
\institute{Université de Mons\\Belgium}
\and
Emmanuel Filiot
\institute{Universit\'e Libre de Bruxelles\\Belgium}
\and
Mickael Randour
\institute{Université de Mons\\Belgium}
\and
Jean-François Raskin
\institute{Universit\'e Libre de Bruxelles\\Belgium}
}
\def\titlerunning{Expectations or Guarantees? I Want It All!}
\def\authorrunning{V. Bruy\`ere, E. Filiot, M. Randour \& J.-F. Raskin}

\begin{document}

\maketitle

\begin{abstract}
When reasoning about the strategic capabilities of an agent, it is important to consider the nature of its adversaries. In the particular context of controller synthesis for quantitative specifications, the usual problem is to devise a strategy for a reactive system which yields some desired performance, taking into account the possible impact of the environment of the system. There are at least two ways to look at this environment. In the classical analysis of two-player quantitative games, the environment is purely antagonistic and the problem is to provide strict performance guarantees. In Markov decision processes, the environment is seen as purely stochastic: the aim is then to optimize the expected payoff, with no guarantee on individual outcomes. 

In this expository work, we report on recent results \cite{bruyere_STACS2014,bruyere_arXiv2013} introducing the beyond worst-case synthesis problem, which is to construct strategies that guarantee some quantitative requirement in the worst-case while providing an higher expected value against a particular stochastic model of the environment given as input. This problem is relevant to produce system controllers that provide nice expected performance in the everyday situation while ensuring a strict (but relaxed) performance threshold even in the event of very bad (while unlikely) circumstances. It has been studied for both the mean-payoff and the shortest path quantitative measures.
\end{abstract}

\section{Introduction}

\smallskip\noindent\textbf{Classical models.} Two-player zero-sum quantitative games~\cite{EM79,ZP96,BCDGR11} and Markov decision processes (MDPs)~\cite{Puterman94,chatterjee_MEMICS11} are two popular formalisms for modeling decision making in adversarial and uncertain environments respectively. In the former, two players compete with opposite goals (zero-sum), and we want strategies for player~1 (the system) that ensure a given \textit{minimal performance against all possible strategies} of player~2 (its environment). In the latter, the system plays against a stochastic model of its environment, and we want strategies that ensure a \textit{good expected overall performance}.  Those two models are well studied and simple optimal memoryless strategies exist for classical objectives such as mean-payoff~\cite{liggett_SR69,EM79,filar1997} or shortest path~\cite{bertsekas_MOR1991,deAlfaro_CONCUR1999}. But both models have clear weaknesses: strategies that are good for the worst-case may exhibit suboptimal behaviors in probable situations while strategies that are good for the expectation may be terrible in some unlikely but possible situations.

\smallskip\noindent\textbf{What if we want both?} In practice, we want strategies that both ensure $(a)$  some worst-case threshold no matter how the adversary behaves (i.e., against any arbitrary strategy) and $(b)$ a good expectation against the expected behavior of the adversary (given as a stochastic model). We study how to construct such finite-memory strategies. We consider finite memory for player~1 as it can be implemented in practice (as opposed to infinite memory). Player~2 is not restricted in his choice of strategies, but we show that simple strategies suffice. Our problem, the \textbf{beyond worst-case synthesis problem}, makes sense for any quantitative measure. We focus on two classical ones: the {\em mean-payoff}, and the {\em shortest path}. Our results are summarized in Table~\ref{summaryTable}.

\renewcommand{\arraystretch}{1.4}
\begin{table}[htb]
\centering
\begin{tabular}{|c|c||c|c|c|}
\cline{3-5} \multicolumn{2}{c|}{} & worst-case & ~expected value~ & \textbf{BWC}\\
\hline
\hline
\multirow{2}{*}{mean-payoff} & ~~complexity~~ & ~~$\NPinter$~~ & $\PTIME$ & \textbf{$\NPinter$}\\
\cline{2-5}
& memory & \multicolumn{2}{c|}{memoryless} & \textbf{pseudo-poly.}\\
\hline
\multirow{2}{*}{~~shortest path~~} & complexity & \multicolumn{2}{c|}{$\PTIME$} & ~~\textbf{pseudo-poly. / $\NPTIME$-hard}~~\\
\cline{2-5}
& memory & \multicolumn{2}{c|}{memoryless} & \textbf{pseudo-poly.}\\
\hline
\end{tabular}
\caption{Overview of decision problem complexities and memory requirements for winning strategies of the first player in games (worst-case), MDPs (expected value) and the BWC setting (combination).}
\label{summaryTable}
\end{table}

\renewcommand{\arraystretch}{1}
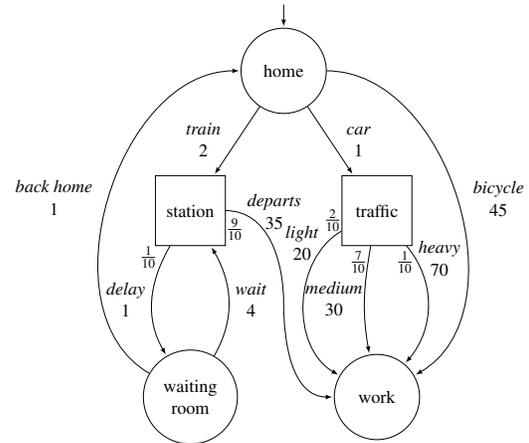
\begin{wrapfigure}{r}{74mm}
  \centering
\scalebox{0.62}{\begin{tikzpicture}[->,>=latex,shorten >=1pt,auto,node
    distance=2.5cm,bend angle=45,font=\normalsize]
    \tikzstyle{p1}=[draw,circle,text centered,minimum size=10mm, text width=15mm]
    \tikzstyle{p2}=[draw,rectangle,text centered,minimum size=15mm]
    \tikzstyle{empty}=[]
    \node[p1] (1) at (0,0) {home};
    \node[p2] (2) at (-2,-3) {station};
    \node[p2] (3) at (2,-3) {traffic};
    \node[p1] (4) at (-2,-7) {waiting room};
    \node[p1] (5) at (2,-7) {work};
    \node[empty] (a) at (-2.9,-4) {$\frac{1}{10}$};
    \node[empty] (b) at (-1.05,-3.4) {$\frac{9}{10}$};
    \node[empty] (c) at (1.05,-3.2) {$\frac{2}{10}$};
    \node[empty] (d) at (1.6,-4.1) {$\frac{7}{10}$};
    \node[empty] (e) at (2.6,-4.1) {$\frac{1}{10}$};
    \coordinate[shift={(0mm,5mm)}] (init) at (1.north);
    \path
    (1) edge node[left] {\begin{tabular}{c} \textit{train}\\2\end{tabular}} (2)
    (1) edge node[right] {\begin{tabular}{c} \textit{car}\\1\end{tabular}} (3) 
    (init) edge (1)
    ;
	\draw[->,>=latex] (4) to[out=150,in=180] node[left] {\begin{tabular}{c} \textit{back home}\\1\end{tabular}} (1);
	\draw[->,>=latex] (1) to[out=0,in=30] node[right] {\begin{tabular}{c} \textit{bicycle}\\45\end{tabular}} (5);
	\draw[->,>=latex] (2) to[out=240,in=120] node[left,xshift=2mm] {\begin{tabular}{c} \textit{delay}\\1\end{tabular}} (4);
	\draw[->,>=latex] (4) to[out=60,in=300] node[right,xshift=-2mm] {\begin{tabular}{c} \textit{wait}\\4\end{tabular}} (2);
	\draw[->,>=latex] (3) to[out=210,in=150] node[left, very near start,xshift=2mm] {\begin{tabular}{c} \textit{light}\\20\end{tabular}} (5);
	\draw[->,>=latex] (3) to[out=260,in=100] node[left,xshift=3mm] {\begin{tabular}{c} \textit{medium}\\30\end{tabular}} (5);
	\draw[->,>=latex] (3) to[out=310,in=50] node[right, very near start,xshift=-3mm] {\begin{tabular}{c} \textit{heavy}\\70\end{tabular}} (5);
	\draw [->,>=latex] (2) to[out=0,in=90] node[right, very near start,xshift=-2mm] {\begin{tabular}{c} \textit{departs}\\35\end{tabular}}(0, -5) to[out=270,in=180] (5);
      \end{tikzpicture}}
      \caption{Player 1 wants to minimize its expected time to reach ``work'', but while ensuring it is less than an hour in all cases.}
\label{fig:exampleTS}
\end{wrapfigure}

\smallskip\noindent\textbf{Example.} Consider the weighted game in Fig.~\ref{fig:exampleTS} to illustrate the {\em shortest path} context. Circle states belong to player~1, square states to player~2, integer labels are durations in minutes, and fractions are probabilities that model the expected behavior of player~2. Player~1 wants a strategy to go from ``home'' to ``work'' such that ``work'' is \textit{guaranteed} to be reached within 60 minutes (to avoid missing an important meeting), and player~1 would also like to minimize the expected time to reach ``work''.

The strategy that minimizes the expectation is to take the car (expectation is 33 minutes) but it is excluded as there is a possibility to arrive after 60 minutes (in case of heavy traffic). Bicycle is safe but the expectation of this solution is 45 minutes. We can do better with the following strategy: try to take the train, if the train is delayed three time consecutively, then go back home and take the bicycle. This strategy is safe as it always reaches ``work'' within 59 minutes and its expectation is $\approx 37,56$ minutes (so better than taking directly the bicycle). Observe that this simple example already shows that, unlike the situation for classical games and MDPs, strategies using memory are strictly more powerful than memoryless ones. Our algorithms are able to decide the existence of (and synthesize) such finite-memory strategies.

\smallskip\noindent\textbf{Related work.} This paper gives an expository presentation of results appeared in \cite{bruyere_STACS2014} (an extended version of the paper can be found in \cite{bruyere_arXiv2013}).

Our problems generalize the corresponding problems for two-player zero-sum games and MDPs. In mean-payoff games, optimal memoryless worst-case strategies exist and the best known algorithm is in $\NPinter$~\cite{EM79,ZP96,BCDGR11}. For shortest path games, where we consider game graphs with strictly positive weights and try to minimize the cost to target, it can be shown that memoryless strategies also suffice, and the problem is in $\PTIME$. In MDPs, optimal expectation strategies are studied in~\cite{Puterman94,filar1997} for both measures: memoryless strategies suffice and they can be computed in~$\PTIME$.

Our strategies are {\em strongly risk averse}: they avoid at all cost outcomes below a given threshold (no matter their probability), and inside the set of those {\em safe} strategies, we maximize expectation. To the best of our knowledge, we are the first to consider such strategies.

Other notions of risk have been studied for MDPs: e.g., in~\cite{WL99}, the authors want to find policies minimizing the probability (risk) that the total discounted rewards do not exceed a specified value; in~\cite{FKR95}, the authors want to achieve a specified value of the long-run limiting average reward at a given probability level (percentile). While those strategies limit risk, they only ensure {\em low probability} for bad behaviors but not their absence, furthermore, they do not ensure good expectation either.

Another body of related work is the study of strategies in MDPs that achieve a trade-off between the expectation and the variance over the outcomes (e.g., \cite{brazdil_LICS2013} for the mean-payoff, \cite{mannor_ICML2011} for the cumulative reward), giving a statistical measure of the stability of the performance. In our setting, we strengthen this requirement by asking for \textit{strict guarantees on individual outcomes}, while maintaining an appropriate expected payoff.

\section{Beyond Worst-Case Synthesis}
\label{sec:preliminaries}

\smallskip\noindent\textbf{Preliminaries.} We consider the classical models of \textit{games} and \textit{MDPs}. Both are based on underlying directed \textit{graphs} with integer weights on edges.

In \textit{games}, the set of vertices, called states, is partitioned between states of the first player, denoted by~$\playerOne$, and states of its adversary, denoted by $\playerTwo$. When the game is in a state belonging to $\player_{i}$, $i \in \lbrace 1, 2\rbrace$, then $\player_{i}$ chooses a successor state according to his \textit{strategy}, which may in general use memory (i.e., depend on the history) and be randomized (i.e., prescribe a probability distribution over successor states). This process gives rise to a \textit{play}, an infinite sequence of states corresponding to a path through the game graph. We assign real values to plays according to a \textit{value function}.

In \textit{MDPs}, the set of states is partitioned between states of $\playerOne$ and stochastic states, where the successor state is chosen according to a given probability distribution. Basically, an MDP is a game where the strategy of $\playerTwo$ is fixed.

When we fix the strategy of $\playerOne$ in an MDP, or the strategies of $\playerOne$ and $\playerTwo$ in a game, we obtain a \textit{Markov chain} (MC), a graph where all successor states are chosen according to a stochastic transition function. Given an MC, it is well-known that measurable sets of plays have uniquely defined probabilities~\cite{vardi_FOCS85}, and if we have a measurable value function, we can also compute the \textit{expected value} or \textit{expectation} of this function when executing the MC from a given initial state.

\smallskip\noindent\textbf{Classical problems.} In games, the {\em worst-case threshold problem} asks if $\playerOne$ has a strategy such that any possible outcome, against any possible strategy of $\playerTwo$, gives a play with a value higher than a given threshold. In MDPs, the {\em expected value threshold problem} asks if $\playerOne$ has a strategy such that the resulting MC yields an expectation higher than a given threshold.

\smallskip\noindent\textbf{Our model.} The {\em beyond worst-case ($\BWC$) problem} asks if $\playerOne$ has a finite-memory strategy ensuring, \textit{simultaneously}, a value greater than a threshold $\thresholdWC$ in the worst-case (i.e., against any strategy of the adversary), and an expected value greater than a threshold $\thresholdExp$ against a given finite-memory stochastic model of the adversary (e.g., representing commonly observed behavior of the environment). The {\em $\BWC$ synthesis problem} asks to synthesize such a strategy if one exists.

\section{Mean-Payoff}
\label{sec:mean_payoff}

\smallskip\noindent\textbf{What was known.} Given a play, its mean-payoff is defined as the (inf or sup) limit of the mean encountered weights along its finite prefixes: essentially, it is the long-run average weight over the infinite play. For the worst-case threshold problem, pure memoryless optimal strategies exist for both players~\cite{liggett_SR69,EM79} and deciding the winner is in $\NPinter$ \cite{ZP96,jurdzinski98,gawlitza2009}. Whether the problem is in $\PTIME$ is a long-standing open problem \cite{BCDGR11,chatterjee_ATVA2013}. Optimal expected values in MDPs can be achieved by memoryless strategies, and the corresponding decision problem can be solved in polynomial time through linear programming~\cite{filar1997}.

\smallskip\noindent\textbf{Our results.} We prove that surprisingly, the BWC problem matches the decision complexity of the simpler worst-case problem, even collapsing to $\PTIME$ if the latter were proved to be in $\PTIME$. Hence, we enrich the modeling and reasoning power over strategies without negative impact on the complexity class.

\begin{theorem}
\label{thm:mp_decisionProblem}
The beyond worst-case problem for the mean-payoff value function is in $\NPinter$ and at least as hard as mean-payoff games.
\end{theorem}

Furthermore, we establish that in contrast to the worst-case and expectation problems, some memory is now needed to win in general. Nevertheless, we show that elegantly implementable strategies suffice, constructed using clever alternation between memoryless strategies based on intuitive counters.

\begin{theorem}
\label{thm:mp_memoryRequirements}
Memory of pseudo-polynomial size may be necessary and is always sufficient to satisfy the BWC problem for the mean-payoff: polynomial in the size of the game and the stochastic model, and polynomial in the weight and threshold values.
\end{theorem}

\smallskip\noindent\textbf{Some key ideas.} Our solving algorithm is too complex to be presented fully in this work. Nonetheless, we here give a few hints of its cornerstones, highlighting crucial aspects of the problem.

\smallskip\noindent\textbf{End-components.} An important part of the algorithm relies on the analysis of \textit{end-components} (ECs) in the MDP, i.e., strongly connected subgraphs in which $\playerOne$ can ensure to stay when playing against the stochastic adversary. This is motivated by two facts. First, under any arbitrary strategy, the set of states that are seen infinitely often along an outcome corresponds with probability one to an EC \cite{courcoubetis_JACM1995,de1997formal}. Second, the mean-payoff function is prefix-independent, therefore the value of any outcome only depends on the states that are seen infinitely often. Hence, the expected mean-payoff that $\playerOne$ can achieve on the MDP depends \textit{uniquely} on the value obtained in the ECs. Inside an EC, we can compute the maximal expected value that can be achieved by $\playerOne$, and this value is the same in all states of the EC \cite{filar1997}.

\smallskip\noindent\textbf{Classification of ECs.} To be efficient w.r.t. the expectated value criterion, an acceptable strategy has to favor reaching ECs with a sufficient expectation, but under the constraint that it also guarantees satisfaction of the worst-case requirement: some ECs with high expected values may still need to be avoided because they do not permit to ensure this constraint. We establish a classification of ECs based on that observation, partitioning them between \textit{winning ECs} (WECs) and \textit{losing ECs} (LECs). Since the total number of ECs may be exponential, providing a representative subclass of polynomial size and computing it efficiently is a crucial point to maintain the overall $\NPinter$ membership.

\smallskip\noindent\textbf{Within a WEC.} We give a particularly interesting family of strategies for $\playerOne$ that both guarantee safe outcomes for the worst-case, and prove to be efficient w.r.t. the expected value. Actually, we establish that the worst-case can be guaranteed \textit{almost for free} in the sense that we can achieve expectations arbitrarily close (but not exactly equal) to what $\playerOne$ could obtain without considering the worst-case requirement at all (i.e., in a classical MDP).

To obtain this result we use a finite-memory \textit{combined strategy}. For two well-chosen parameters $\stepsExp, \stepsWC \in \nat$, it is informally defined as follows: in phase $\typeA$, play a memoryless expected value optimal strategy for $\stepsExp$ steps and memorize $\cmbSum \in \integ$, the sum of weights along these steps; in phase $\typeB$, if $\cmbSum > 0$, go to~$\typeA$, otherwise play a memoryless worst-case optimal strategy for $\stepsWC$ steps, then go to $\typeA$. In phases $\typeA$, $\playerOne$ tries to increase its expectation and approach its optimal one, while in phase $\typeB$, he compensates, if needed, losses that occurred in phase $\typeA$.

The crux of the proof is to establish that adequate values of the parameters $\stepsExp$ and $\stepsWC$ exist. Essentially, $\stepsExp$ needs to be big enough so that the overall expectation is close to the optimal, but then $\stepsWC$ also needs to grow to be able to compensate sufficiently for the worst-case, hence lowering to some extent the overall expectation. Using results related to Chernoff bounds and Hoeffding's inequality in MCs \cite{tracol_ORL2009,glynn_SPL2002}, we are able to show that the probability of having to compensate decreases exponentially when $\stepsExp$ increases, while $\stepsWC$ only needs to be polynomial in $\stepsExp$. Overall, this implies the desired result that the parameters can be taken large enough for the strategy to be $\varepsilon$-optimal w.r.t. the expectation while worst-case safe.

\section{Shortest Path}
\label{sec:shortest_path}

\smallskip\noindent\textbf{What was known.} In this context, we consider game graphs where all weights are strictly positive, and a target set of states that $\playerOne$ wants to reach while giving an upper bound on the cost to reach it. Hence the inequalities of the BWC problem are reversed. Given a play, the value function for the shortest path computes the sum of weights up to the first encounter of a state belonging to the target set, or assigning infinity if the play never reaches such a state. The worst-case threshold problem takes polynomial time, as a winning strategy of $\playerOne$ should avoid all cycles (because they yield strictly positive costs), hence usage of attractors and comparison of the worst possible sum of costs with the threshold suffices. For the expected value threshold problem, memoryless strategies suffice and the problem is in $\PTIME$ \cite{bertsekas_MOR1991,deAlfaro_CONCUR1999}.

\smallskip\noindent\textbf{Our results.} In contrast to the mean-payoff case where we could maintain the complexity of the worst-case problem, we here provide an algorithm which operates in pseudo-polynomial time instead of truly-polynomial time. Nevertheless, we prove that the problem is actually $\NPTIME$-hard (reduction from the \textit{$K^{th}$ largest subset problem} \cite{garey_FNY1979}), hence establishing that a truly-polynomial-time algorithm is highly unlikely.

 \begin{theorem}
 \label{thm:ts_pseudoPoly}
The beyond worst-case problem for the shortest path can be solved in pseudo-polynomial time: polynomial in the size of the underlying game graph, the stochastic model of the adversary and the encoding of the expected value threshold, and polynomial in the value of the worst-case threshold. The beyond worst-case problem for the shortest path is $\NPTIME$-hard.
 \end{theorem}

Once again, we show that pseudo-polynomial memory is both necessary and sufficient. Recall that the example of Fig. \ref{fig:exampleTS} already required memory to achieve some thresholds pair for the BWC problem.

\begin{theorem}
\label{thm:ts_memory}
Memory of pseudo-polynomial size may be necessary and is always sufficient to satisfy the BWC problem for the shortest path: polynomial in the size of the game and the stochastic model, and polynomial in the worst-case threshold value.
\end{theorem}

\smallskip\noindent\textbf{Some key ideas.} The shortest path setting has a useful property: the set of all winning strategies of~$\playerOne$ for the worst-case threshold problem can be represented through a finite game. Indeed, we construct, from the original game $\game$ and the worst-case threshold $\thresholdWC$, a new game $\gameTS$ such that there is a bijection between the strategies of $\playerOne$ in $\gameTS$ and the strategies of $\playerOne$ in the original game~$\game$ that are winning for the worst-case requirement: we unfold the original graph, tracking the current value of the sum of weights \textit{up to the threshold $\thresholdWC$}, and integrating this value in the states of an expanded graph. In the corresponding game $\game'$, we compute the set of states $R$ from which~$\playerOne$ can reach the target set with cost lower than~$\thresholdWC$ and we define the subgame $\gameTS = \game' \reduc R$ such that any path in $\gameTS$ satisfies the worst-case requirement.

Assuming that $\gameTS$ is not empty, we can now combine it with the stochastic model of the adversary to construct an MDP in which we search for a $\playerOne$ strategy that ensures reachability of the target set with an expected cost lower than the expectation threshold. If it exists, it is guaranteed that it will also satisfy the worst-case requirement against any strategy of $\playerTwo$ thanks to the bijection evoked earlier.

Hence, in the case of the shortest path, our approach is sequential, first solving the worst-case, then optimizing the expected value among the worst-case winning strategies. 
This sequential algorithm is depicted through Example~\ref{ex}.
Observe that such an approach is not applicable to the mean-payoff, as in that case there exists no obvious finite representation of the worst-case winning strategies.
\begin{example}
\label{ex}
Consider the game $\game$ depicted in Fig.~\ref{fig:bwc_sp_game}. We want to synthesize a BWC strategy of $\playerOne$ that minimizes the expected cost up to the target set $\{\state_{3}\}$ under the (strict) worst-case threshold $\thresholdWC = 8$.

\begin{figure}[htb]
  \centering   
 \scalebox{0.7}{\begin{tikzpicture}[->,>=latex,shorten >=1pt,auto,node
    distance=2.5cm,bend angle=45,font=\Large]
    \tikzstyle{p1}=[draw,circle,text centered,minimum size=10mm]
    \tikzstyle{p2}=[draw,rectangle,text centered,minimum size=10mm]
    \tikzstyle{empty}=[]
    \node[p1] (1) at (0,0) {$\state_{1}$};
    \node[p2] (2) at (4,0) {$\state_{2}$};
    \node[p1,double] (3) at (2,-3) {$\state_{3}$};
    \node[empty] (a) at (3.5,0.9) {$\frac{1}{2}$};
    \node[empty] (b) at (3.8,-0.9) {$\frac{1}{2}$};
    \coordinate[shift={(-5mm,0mm)}] (init) at (1.west);
    \path
    (1) edge node[below] {$1$} (2)
    (2) edge node[right,xshift=2mm] {$1$} (3)
    (1) edge node[left, xshift=-2mm] {$5$} (3)
    (init) edge (1)
    ;
	\draw[->,>=latex] (2) to[out=140,in=40] node[above] {$1$} (1);
      \end{tikzpicture}}
      \caption[Simple BWC shortest path game]{Simple BWC shortest path game with target set $\{\state_{3}\}$ and worst-case threshold $\thresholdWC = 8$.}
\label{fig:bwc_sp_game}
\end{figure}
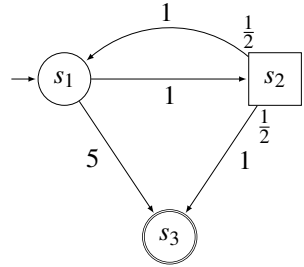

First, we unfold this game $\game$ up to the worst-case threshold (excluded), and obtain the game $\game'$ represented in Fig.~\ref{fig:bwc_sp_unfolding}. Observe that as soon as the worst-case threshold is reached, we stop the unfolding and associate symbol $\tsFailSymbol$: the worst-case requirement is lost if such states are reached. This guarantees a finite (and at most pseudo-polynomial size) unfolding.

\begin{figure}[htb]
  \centering   
\scalebox{0.46}{\begin{tikzpicture}[->,>=latex,shorten >=1pt,auto,node
    distance=2.5cm,bend angle=45,font=\LARGE]
    \tikzstyle{p1}=[draw,circle,text centered,minimum size=16mm]
    \tikzstyle{p2}=[draw,rectangle,text centered,minimum size=16mm]
    \tikzstyle{empty}=[]
    \node[p1] (1) at (0,0) {$\state_{1}, 0$};
    \node[p2] (2) at (4,0) {$\state_{2}, 1$};
    \node[p1] (3) at (8,0) {$\state_{1}, 2$};
    \node[p2] (4) at (12,0) {$\state_{2}, 3$};
    \node[p1] (11) at (16,0) {$\state_{1}, 4$};
    \node[p2] (12) at (20,0) {$\state_{2}, 5$};
    \node[p1] (13) at (24,0) {$\state_{1}, 6$};
    \node[p2] (14) at (28,0) {$\state_{2}, 7$};
    \node[p1] (15) at (32,0) {$\state_{1}, \tsFailSymbol$};
    \node[p1,double,thick] (7) at (4,-5) {$\state_{3}, 2$};
    \node[p1,double,thick] (8) at (12,-5) {$\state_{3}, 4$};
    \node[p1] (9) at (16,-5) {$\state_{3}, \tsFailSymbol$};
    \node[p1,double,thick] (10) at (20,-5) {$\state_{3}, 6$};
    \node[p1,double,thick] (18) at (0,-5) {$\state_{3}, 5$};
    \node[p1,double,thick] (19) at (8,-5) {$\state_{3}, 7$};
    \node[empty] (a) at (3.8,-1.4) {$\frac{1}{2}$};
    \node[empty] (d) at (5.1,0.5) {$\frac{1}{2}$};
    \node[empty] (b) at (11.8,-1.4) {$\frac{1}{2}$};
    \node[empty] (e) at (13.1,0.5) {$\frac{1}{2}$};
    \node[empty] (c) at (19.8,-1.4) {$\frac{1}{2}$};
    \node[empty] (f) at (21.1,0.5) {$\frac{1}{2}$};
    \node[empty] (c) at (27.7,-1.4) {$\frac{1}{2}$};
    \node[empty] (f) at (29.1,0.5) {$\frac{1}{2}$};
    \coordinate[shift={(-5mm,0mm)}] (init) at (1.west);
    \path
    (1) edge node[right] {$5$} (18)
    (init) edge (1);
    \path (1) edge node[above] {$1$} (2);
    \path (1) edge [ultra thick] node[above] {$1$} (2);
    \path (2) edge node[right] {$1$} (7)
    (2) edge node[above] {$1$} (3);
    \path (3) edge node[right] {$5$} (19);
    \path (3) edge [ultra thick] node[right] {$5$} (19);
    \path (3) edge node[above] {$1$} (4);
    \path (4) edge node[above] {$1$} (11)
    (4) edge node[right] {$1$} (8);
    \path (11) edge node[above] {$1$} (12)
    (11) edge node[right] {$5$} (9);
    \path (12) edge node[above] {$1$} (13)
    (12) edge node[right] {$1$} (10);
    \path (13) edge node[above] {$1$} (14)   
    (14) edge node[above] {$1$} (15);
	
    \draw[->,>=latex] (13) to[out=280,in=310] node[right,xshift=3mm] {$5$} (9);
	\draw[->,>=latex] (14) to[out=270,in=300] node[right,xshift=4mm] {$1$} (9);
      \end{tikzpicture}}
      \vspace{-1cm}
      \caption[Unfolding of the game of Fig.~\ref{fig:bwc_sp_game}]{Unfolding of the game of Fig.~\ref{fig:bwc_sp_game}: worst-case winning requires to reach a double state. Thick edges represent the strategy that minimizes the expected cost while ensuring this worst-case.}
\label{fig:bwc_sp_unfolding}
\end{figure}
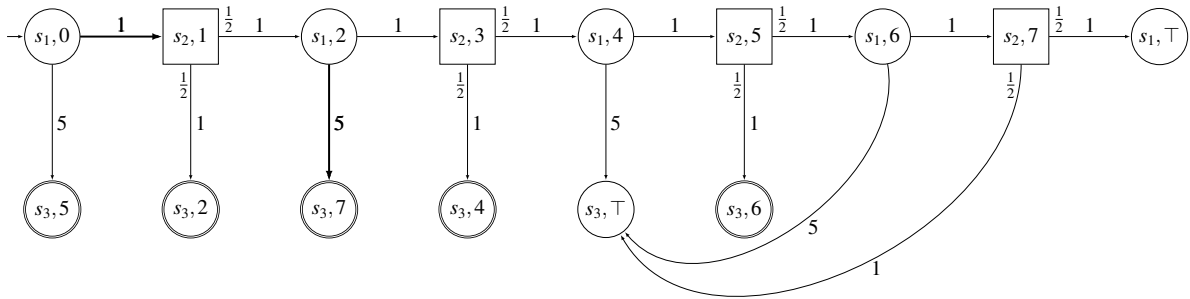

Therefore, it is clear that a BWC strategy of $\playerOne$ must ensure reachability of states of $\game'$ that represent reaching the target state with a total cost strictly less than the worst-case threshold. Those states are depicted by double circles in the figure. Hence, $\playerOne$ must stay within the attractor of those double states. It implies that state $(\state_{2}, 3)$ of the unfolding and subsequent states are off-limits.

Knowing that, it now suffices to minimize the expected value within the safe region, which is achieved by the memoryless (with regard to $\game'$) strategy that chooses to go in $(\state_{2}, 1)$ from $(\state_{1}, 0)$ and to $(\state_{3}, 7)$ from $(\state_{1}, 2)$. This strategy is depicted by the thick edges on the figure. Observe that this strategy is memoryless in $\game'$, hence requires at most pseudo-polynomial memory in $\game$.
\end{example}

\section{Future Work}
We believe that the beyond worst-case framework is a powerful one, well-suited for specifications combining the quest of high expected performance with the need for strong worst-case guarantees. We want to build on the results presented here and consider several extensions of the initial setting.

The first line of work is applying the problem to other well-known quantitative measures and to more general classes of games (for example decidable classes of games with imperfect information~\cite{degorre_CSL2010,DBLP:journals/corr/HunterPR13}).

A second interesting question is the extension of our results for mean-payoff and shortest path to multi-dimension games. It is already known that multi-dimension games are more complex than one-dimension ones for the worst-case threshold problem alone~\cite{chatterjee_FSTTCS10,chatterjee_CONCUR2012}. Hence, a leap in complexity is also to be expected for the beyond worst-case problem.

Given the relevance of the framework for practical applications, it would certainly be worthwhile to develop tool suites supporting it. We could for example build on symblicit implementations recently developed for monotonic Markov decision processes by Bohy et al.~\cite{2014arXiv1402.1076B}.

Links outside computer science are also of interest. Economics is interested in strategies (i.e., investor profiles) that ensure both sufficient risk-avoidance and profitable expected return. Mathematical models powerful enough to tackle the previously discussed problems could be an advantage. A related approach to such questions is the concept of \textit{solvency games} introduced by Berger et al.~\cite{DBLP:conf/fsttcs/BergerKSV08}, and extended by Br\'azdil et al.~\cite{DBLP:conf/fsttcs/BrazdilCFNS13}. Solvency games provide a framework for the analysis of risk-averse investors trying to avoid bankruptcy.

\bibliographystyle{eptcs}
\bibliography{bwc_bib}

\end{document}